\documentstyle[preprint,amsfonts,amssymb,aps,12pt]{revtex}
\newcommand{\mbf}[1]{\mbox{\boldmath$#1$\unboldmath}}
\newcommand{\D}{{\rm d}}
\renewcommand{\mathrm}[1]{{\rm #1}}
\input{epsf}
\tighten
\begin{document}
\draft
\title{Molecular Dynamics Results on the Pressure Tensor\\of Polymer Films}
\author{F. Varnik$^1$\footnote{To whom correspondence should be addressed.
Email: {\sf varnik@plato.physik.uni-mainz.de}}, J. Baschnagel$^2$, K. Binder$^1$\\[2mm]}
\address{$^1$Institut f\"ur Physik, Johannes-Gutenberg Universit\"at, 55099 Mainz, Germany\\[1mm]}
\address{$^2$Institut Charles Sadron, 6 rue Boussingault, 67083 Strasbourg Cedex, France}
\maketitle

\begin{abstract}
Polymeric thin films of various thicknesses, confined between two repulsive walls,
have been studied by molecular dynamics simulations. Using the anisotropy of the 
perpendicular, $P_\mathrm{N}(z)$, and parallel components, $P_\mathrm{T}(z)$, of the 
pressure tensor the surface tension of the system is calculated for a wide range 
of temperature and for various film thicknesses. Three methods of determining the 
pressure tensor are compared: the method of Irving and Kirkwood (IK), an approximation 
thereof (IK1), and the method of Harasima (H). The IK- and the H-methods differ
in the expression for $P_{\mathrm {T}}(z)$ ($z$ denotes the
distance from the wall), but yield the same formula for the normal component 
$P_{\mathrm {N}}(z)$. When evaluated by MD (or MC)-simulations $P_{\mathrm {N}}(z)$
is constant, as required by mechanical stability. Contrary to that, the IK1-method 
leads to strong oscillations of $P_{\mathrm {N}}(z)$. However, all methods give 
the same expression for the total pressure when integrated over the whole system, and
thus the same surface tension, whereas the so-called surface of tension, $z_\mathrm{s}$, 
depends on the applied method. The difference is small for the IK- and H-methods, while 
the IK1-method leads to values that are in conflict with the interpretation of $z_\mathrm{s}$ 
as the effective position of the interface.\\[2mm]
\end{abstract}

\section{Introduction}
The aim of statistical mechanics is to relate macroscopic quantities to
microscopic degrees of freedom. An example for this connection 
is the virial equation of the pressure. Consider a system of volume $V$ with
$M$ particles which interact by a pair potential $U$. Let the distance 
between two particles be denoted $\mbf{R}$ ($R= |\mbf{R}|$). The pressure can 
then be written as a sum of two parts,
\begin{equation}
p= k_\mathrm{B}T \rho - \frac{1}{6} \int R\frac{\D U(R)}{\D R}
\rho^{(2)}(R) \,\D^3\mbf{R} \; ,
\label{eq:virial}
\end{equation}
a kinetic (ideal-gas) part $k_\mathrm{B}T \rho$ ($\rho=M/V$), which arises from the average 
kinetic energy and the momentum transfer of the particles on the container walls,
and a potential part which accounts for the intermolecular forces. Two particles experience 
an interaction force $-\mbf{R}U^\prime(R)/R$. When weighing the corresponding virial,
$-RU^\prime(R)$, with the average density, $\rho^{(2)}(R)$, of a particle at distance $R$ 
from another one and integrating over all possible separations, one obtains
the contribution of the potential
to the pressure. There are different ways to derive Eq.~(\ref{eq:virial}) (see
\cite{Hansen,RowlinsonWidom,Haile}, for instance), but none of these routes can readily
be generalized to inhomogeneous systems. They all use the isotropy of space somewhere 
in the derivation and take $p$ as a scalar.

In inhomogeneous systems, however, the pressure in general depends on the spatial
direction and on the position $\mbf{r}$ where it is determined: It is a tensor 
${\mathbf{P}}(\mbf{r})$. Nonetheless, the pressure tensor can still be split into a
kinetic part, ${\mathbf{P}}^{K}$, and a potential part, ${\mathbf{P}}^{U}$:
\begin{equation}
{\mathbf{P}}(\mbf{r}) = {\mathbf{P}}^{K}(\mbf{r}) + {\mathbf{P}}^{U}(\mbf{r}) \;.
\label{eq:split}
\end{equation}
The kinetic part may be expressed by a generalization of the
ideal-gas contribution,
\begin{equation}
{\mathbf{P}}^{K}(\mbf{r}) =k_{\mathrm B}T \rho(\mbf{r})\hat{\mathbf 1}\;,
\label{eq:kin}
\end{equation}
where $\rho(\mbf{r})$ is the density at $\mbf{r}$ and $\hat{\mathbf {1}}$ a
$3\times 3$ unit matrix. 

On the other hand, there seems to be no unique expression
for ${\mathbf{P}}^{U}(\mbf{r})$ \cite{RowlinsonWidom,IK,Harasima,SchoHend,Lovett,ToddEvans,%
ToddEvans2,WTR}. 
The origin of this problem may be explained as follows: The pressure tensor can be
defined by the infinitesimal force ${\mathrm d}\mbf{F}$ acting across an 
infinitesimal surface ${\mathrm d}\mbf{A}$ which is located at $\mbf{r}$:
\begin{equation}
{\mathrm d}\mbf{F}(\mbf{r})=-{\mathrm d}\mbf{A}\cdot {\mathbf P}(\mbf{r})\;.
\label{eq:defP}
\end{equation}
If a particle moves across ${\mathrm d}\mbf{A}$, the resulting momentum transfer
contributes to ${\mathbf{P}}^{K}(\mbf{r})$. Since the momentum is associated with the 
particle position, it is a single particle property which may be well localized in space
(see however \cite{pt1_of_referee}). The ambiguity in the calculation of ${\mathbf P}(\mbf{r})$
arises from the interaction between two particles: Which particles should contribute to the 
force at $\mbf{r}$? Somehow the non-local two-particle force, $-U^\prime(R)$, has to be
reduced to a local force ${\mathrm d}\mbf{F}(\mbf{r})$ \cite{Lovett}. This ambiguity was
already pointed out in the seminal work of Irving and Kirkwood, and they required 
that ``all definitions must have this in common -- that the stress between a pair of 
molecules be concentrated near the line of centers. When averaging over a domain 
large compared with the range of intermolecular force, these differences are washed 
out, and the ambiguity remaining in the macroscopic stress tensor ($\!\!$\cite{footnote})
is of negligible order'' (footnote on p.~829 of \cite{IK}). 

In the present paper, we apply common ways to calculate ${\mathbf{P}}^{U}
(\mbf{r})$ to a model of a glassy polymer film and determine the surface
tension as a function of temperature. This work serves as a preparation for 
simulations on the sluggish relaxation of the film in the supercooled
state \cite{vbb_confit}. The paper is organized as follows: In Sect.~\ref{sec:theo}
we discuss the theoretical background of various approaches to ${\mathbf{P}}^{U}
(\mbf{r})$. Section~\ref{sec:sim} presents details of the model and simulation 
technique, and Sect.~\ref{sec:results} compiles the results. The final section 
contains our conclusions.
\section{Theoretical Background}
\label{sec:theo}
\subsection{The Methods of Irving and Kirkwood and of Harasima}
\label{subsec:ikh}
\begin{figure}[pbt]
\unitlength=1mm
\begin{picture}(150,80)
\put(3,10){
\epsfxsize=150mm
\epsffile{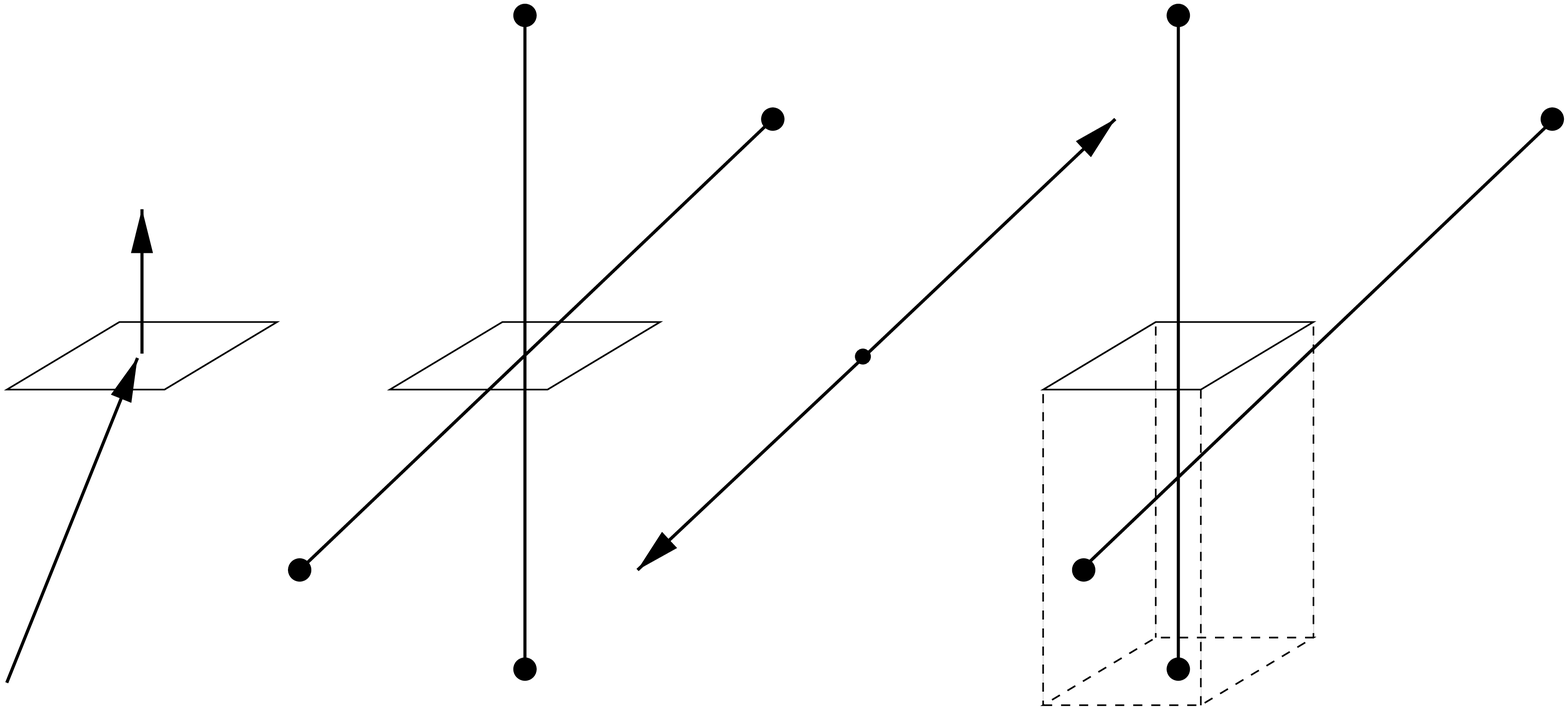}
\put(-139,49){\makebox(0,0)[lb]{\mbox{$\D \mbf{A}$}}}
\put(-148,17){\makebox(0,0)[lb]{\mbox{$\mbf{r}$}}}
\put(-119.5,10){\makebox(0,0)[lb]{\mbox{$\mbf{r}_1$}}}
\put(-82.5,56){\makebox(0,0)[lb]{\mbox{$\mbf{r}_2$}}}
\put(-105.5,1){\makebox(0,0)[lb]{\mbox{$\mbf{r}_1^{\prime}$}}}
\put(-97.5,65){\makebox(0,0)[lb]{\mbox{$\mbf{r}_2^{\prime}$}}}
\put(-87,10){\makebox(0,0)[lb]{\mbox{$\mbf{r}_1$}}}
\put(-51,56){\makebox(0,0)[lb]{\mbox{$\mbf{r}_2$}}}
\put(-76,21){\makebox(0,0)[lb]{\mbox{$-\alpha \mbf{R}$}}}
\put(-76,43){\makebox(0,0)[lb]{\mbox{$(1-\alpha)\mbf{R}$}}}
\put(-66,30){\makebox(0,0)[lb]{\mbox{$\mbf{r}$}}}
%
\put(-44,10){\makebox(0,0)[lb]{\mbox{$\mbf{r}_3$}}}
\put(-8,56){\makebox(0,0)[lb]{\mbox{$\mbf{r}_4$}}}
\put(-42.5,1){\makebox(0,0)[lb]{\mbox{$\mbf{r}_1^{\prime}$}}}
\put(-35,65){\makebox(0,0)[lb]{\mbox{$\mbf{r}_2^{\prime}$}}}
\put(-135,1){\makebox(0,0)[lb]{\mbox{(a)}}}
\put(-70,1){\makebox(0,0)[lb]{\mbox{(b)}}}
\put(-15,1){\makebox(0,0)[lb]{\mbox{(c)}}}
}
\end{picture}
\caption[]{
Schematic illustration of the different contributions to ${\mathbf{P}}^U(\mbf{r})$ 
which are taken into account by Irving and Kirkwood (IK-method) and by Harasima
(H-method). Let $\D \mbf{A}$ be an infinitesimal surface situated at position 
$\mbf{r}$ [panel (a)]. In the IK-method all particles whose center line passes
through $\D \mbf{A}$ contribute to the force felt across the surface [panel (b)],
whereas Harasima assumes that the interaction between the particles inside a 
prisma with base $\D A$ and those on the side to which $\D \mbf{A}$ is 
pointing causes the force at $\mbf{r}$ [panel (c)]. Panel (b) shows two possible
contributions in the IK-method. If $\mbf{R}=\mbf{r}_2-\mbf{r}_1$, the position
vectors of the particles can also be expressed as: $\mbf{r}_1=\mbf{r}-\alpha
\mbf{R}$ and  $\mbf{r}_2=\mbf{r}+(1-\alpha)\mbf{R}$ ($0\leq \alpha \leq 1$)
[see Eq.~(\ref{eq:IK:A6})]. The interaction between $\mbf{r}_1^\prime$ and 
$\mbf{r}_2^\prime$ is also taken into account in the H-method, but not that between
$\mbf{r}_1$ and $\mbf{r}_2$. On the other hand, particles at $\mbf{r}_3$ and
$\mbf{r}_4$ ($=\mbf{r}_3+\mbf{R}$) contribute in Harasima's approach, whereas 
they don't in the IK-method.
}
\label{fig:ik+h}
\end{figure}

Irving and Kirkwood~\cite{IK} gave a definition of the ${\mathbf P}^{U}$-tensor by
starting from a statistical mechanical derivation of the equations of hydrodynamics
and by making a special choice for the particles that contribute to the local force:
Only those pairs of particles should give rise to $\D\mbf{F}(\mbf{r})$
for which the line connecting their centers of mass passes through the infinitesimal 
surface $\D\mbf{A}$ (see Fig.~\ref{fig:ik+h}) \cite{RowlinsonWidom}. With this choice 
they obtained the following expression for the potential part of the pressure tensor
\begin{equation}
{\mathbf{P}}^{U} ( {\mbf{r}} ) = -\frac{1}{2}
\int \frac{{\mbf{R}}{\mbf{R}}}{R} \,\, U'(R)
\left( 
\int_{0}^{1} \D\alpha\,\, 
\rho^{(2)} ( {\mbf{r}} - \alpha{\mbf{R}};
{\mbf{r}} + (1-\alpha) {\mbf{R}} ) 
\right)
\D^3 {\mbf{R}} \;, 
\label{eq:IK:A6}
\end{equation}
where ${\mbf{R}}{\mbf{R}}$ is a dyadic, $U'(R)=\D U/\D R$, and $\rho^{(2)}(\mbf{r};
\mbf{r}^\prime)$ denotes the two-particle density
\begin{equation}
\rho^{(2)} ( {\mbf{r}}; {\mbf{r}'} ) = \Big \langle  \sum_{i\neq j}
\delta( {\mbf r}_{i}-{\mbf r} ) \,
\delta( {\mbf r}_{j}-{\mbf r}' ) \Big \rangle \; .   
\label{def:rho2}
\end{equation}
Using Eq.~(\ref{def:rho2}) one obtains from~(\ref{eq:IK:A6}) 
\begin{equation}
{\mathbf P}^{U} ( {\mbf r} ) = -\frac{1}{2}  \Big \langle  \sum_{i \neq j}
\frac{\mbf{r}_{ij}\mbf{r}_{ij}}{r_{ij}} \,\, U'(r_{ij})
\int_{0}^{1} \D\alpha\,\, \delta({\mbf r}_{i}-{\mbf r}+\alpha\, {\mbf r}_{ij}) 
\Big \rangle \;,
\label{eq:IK:A6'}
\end{equation}
where  $\mbf{r}_{ij}=\mbf{r}_{j}-\mbf{r}_{i}$ (${r}_{ij}=|\mbf{r}_{ij}|$).

Equation~(\ref{eq:IK:A6}) can be interpreted as follows: The term $-\mbf{R}\mbf{R}
U^\prime/R$ is a tensorial generalization of the virial $-RU^\prime$ of the integrand
in Eq.~(\ref{eq:virial}). It accounts for the force $\mbf{R}U^\prime/R$ that a particle 
at $\mbf{r}_1$ experiences from another particle at $\mbf{r}_2$
($\mbf{R}=\mbf{r}_2 - \mbf{r}_1$). The virial has to be multiplied by 
the probability of finding two particles at $\mbf{r}_1$ and $\mbf{r}_2$. The
probability is proportional to the density $\rho^{(2)}( \mbf{r}_1;\mbf{r}_2)$ which 
depends explicitly on both particle positions for inhomogeneous systems. Therefore, 
different values of $\rho^{(2)}(\mbf{r}_1;\mbf{r}_2)$ are obtained for fixed $\mbf{R}$ 
when shifting particle 1 or 2 to position $\mbf{r}$, where the pressure shall be 
determined, i.e., for $\mbf{r}_1=\mbf{r}$ ($\alpha=0$) or $\mbf{r}_2=\mbf{r}$ 
($\alpha=1$) (see Fig.~\ref{fig:ik+h}). The integral over $\alpha$ takes all 
of these contributions into account. The outer integral finally sums over the possible
vectors $\mbf{R}$ which pass through $\D \mbf{A}$.
Equations~(\ref{eq:IK:A6}) and (\ref{eq:IK:A6'}) are general and apply to systems of
any shape if the particles interact by a pair potential. In the following we are
interested in thin (polymer) films confined between two impenetrable walls.
For systems with planar geometry the pressure tensor, ${\mathbf{P}}$, depends 
only on the distance, $z$, from the wall \cite{RowlinsonWidom,WTR}.
Furthermore, the non-diagonal components of ${\mathbf{P}}$ vanish in 
thermal equilibrium and it can be written
as (see Sect.~\ref{subsec:mechanical_stability})
\begin{equation}
{\mathbf{P}}(z)=({\mbf{e}}_x{\mbf{e}}_x + {\mbf{e}}_y{\mbf{e}}_y)P_\mathrm{T}(z)+
{\mbf{e}}_z {\mbf{e}}_z P_\mathrm{N}(z) \;,
\label{eq:PTpg}
\end{equation}
where $\mbf{e}_x$, $\mbf{e}_y$, $\mbf{e}_z$ are orthogonal unit vectors and the
lateral, $P_\mathrm{T}(z)$, and normal component, $P_\mathrm{N}(z)$, of ${\mathbf{P}}(z)$
are given by
\begin{equation}
P_{zz}(z)=P_\mathrm{N}(z) \quad \mbox{and} \quad
P_{xx}(z)=P_{yy}(z)=P_\mathrm{T}(z) \; .
\label{eq:PnPt}
\end{equation}
Using
\[
\int_0^1\D \alpha\, \delta(z-\alpha z_{ij} -z_i) 
= \frac{1}{|z_{ij}|} \Theta \bigg(\frac{z-z_i}{z_{ij}}\bigg)
\Theta \bigg(\frac{z_j-z}{z_{ij}}\bigg) \;,
\]
and averaging Eq.~(\ref{eq:IK:A6'}) over the tangential coordinates
one obtains~\cite{WTR,RaoBerne}
\begin{eqnarray}
{\mathbf P}^{U} (z) &=& \frac{1}{A} \int\!\!\int {\mathbf P}^{U} ( {\mbf r} )
\, \D x\, \D y \nonumber\\
&=& -\frac{1}{2A} \bigg \langle \sum_{i\neq j}\frac{\mbf{r}_{ij}\mbf{r}_{ij}}{r_{ij}}\,\, U'(r_{ij})
\, \frac{1}{|z_{ij}|}\Theta\bigg(\frac{z-z_i}{z_{ij}}\bigg)\,\,\Theta\bigg(\frac{z_j-z}{z_{ij}}
\bigg) \bigg\rangle \;,
\label{eq:IK:film}
\end{eqnarray}
where $A$ is the area of a plane in tangential direction. With Eq.~(\ref{eq:IK:film})
this leads to the following (full) expressions for the normal
and tangential components of the pressure tensor for planar systems (IK-method)
\begin{eqnarray}
P_{\mathrm {N}}^{\mathrm{IK}}(z)&\!\!\!=\!\!\!&\rho(z)k_{\mathrm B}T\!-\!\frac{1}{2A} \bigg \langle
\sum_{i\neq j} \frac{|z_{ij}|}{r_{ij}} U'(r_{ij}) \Theta \Big(  \frac{z-z_i}{z_{ij}} \Big) 
\Theta \Big( \frac{z_j-z}{z_{ij}} \Big)  \bigg\rangle \;,
\label{eq:p_N:RaoBerne}\\
P_{\mathrm {T}}^{\mathrm{IK}}(z)&\!\!\!=\!\!\!&\rho(z)k_{\mathrm B}T\!-\!\frac{1}{4A}  \bigg \langle
\sum_{i\neq j} 
\frac{{x}^2_{ij}+{y}^2_{ij}}{r_{ij}} \frac{U'(r_{ij})}{|z_{ij}|} 
\Theta \Big(  \frac{z-z_i}{z_{ij}} \Big) \Theta \Big( \frac{z_j-z}{z_{ij}} \Big)\bigg  \rangle \;,
\label{eq:p_T:RaoBerne}
\end{eqnarray}
where $\rho(z)$ denotes the density at $z$ averaged over tangential coordinates $x$ and $y$. These
equations are valid only in thermal equilibrium (for an extension to non-equilibrium situations
see \cite{ToddEvans,ToddEvans2}).

In addition to the IK-expressions the formulas of Harasima are often used in the literature
\cite{RowlinsonWidom,Harasima}. They are obtained from a different choice of the contributing 
interactions (see Fig.~\ref{fig:ik+h}): Harasima considered a prisma whose base is $\D A$. 
The force $\D \mbf{F}(\mbf{r})$ 
is thought to result from all interactions between particles in the prisma and those on the 
side of $\D A$ to which the vector $\D \mbf{A}$ points. This also includes particles
whose center line does not pass through $\D A$.  Harasima's choice corresponds to a contour which 
goes parallel to the walls (or the planar surface) from $\mbf{r}_1$ to $(x_2, y_2,z_1)$
and then along the normal to  $\mbf{r}_2$~\cite{RowlinsonWidom,WTR}.
Using this convention he obtained the same results for the normal component as
Irving and Kirkwood [Eq.~(\ref{eq:p_N:RaoBerne})],
\begin{equation}
P_{\mathrm {N}}^{\mathrm{H}}(z)=P_{\mathrm {N}}^{\mathrm{IK}}(z) \;,
\label{eq:pN:H}
\end{equation}
but a different expression for the lateral component of the pressure 
tensor~\cite{RowlinsonWidom,Harasima}
\begin{eqnarray}
P^{\mathrm{H}}_{\mathrm {T}}(z) &=&\rho(z)k_{\mathrm B}T-\frac{1}{4A} \bigg \langle
\sum_{i\neq j}	\frac{{x}^2_{ij}+{y}^2_{ij}}{r_{ij}} 
	U'(r_{ij})\, \delta(z_i-z) \bigg \rangle \;.
\label{eq:p_T:Harasima}
\end{eqnarray}
Thus, the \emph{tangential component}, $P_{\mathrm {T}}$, of the pressure 
tensor is not uniquely defined. Consequently, the \emph{pressure anisotropy},
$P_{\mathrm {N}}-P_{\mathrm {T}}$, is ambiguous. This ambiguity is extensively 
discussed in the literature \cite{RowlinsonWidom,IK,Harasima,SchoHend,Lovett,ToddEvans,%
ToddEvans2,WTR,RaoBerne}.

However, the integral over $z$ of Eq.~(\ref{eq:p_T:RaoBerne}) is identical to that of
Eq.~(\ref{eq:p_T:Harasima}). This implies that both the IK and the H-methods yield the same 
results for any physical quantity which does not depend on the local profile of the pressure 
tensor.  In particular, they lead to the same values of the surface tension $\gamma$ 
(Kirkwood--Buff formula \cite{RowlinsonWidom})
\begin{eqnarray}
2\gamma 
&=&
\int_{-D/2}^{+D/2} \Big[P_{\mathrm {N}}(z)-P_{\mathrm {T}}(z) \Big]\D z 
\label{eq:def:gamma} \\
&=&
\frac{1}{4A} \Bigg \langle \sum_{i\neq j} \frac{{r}^2_{ij}-3{z}^2_{ij}}{r_{ij}} 
U'(r_{ij})\Bigg \rangle \;.
\label{eq:kb:gamma}
\end{eqnarray}
The factor $2$ arises from the existence of two walls at $z=-D/2$ and $z=D/2$ in our
simulation, $D$ being the distance from one wall to the other (i.e., the film thickness).
However, moments of $P_{\mathrm {N}}-P_{\mathrm {T}}$, such as the so-called
``surface of tension'' $z_\mathrm{s}$, i.e., the position where the surface tension
acts,
\begin{equation}
z_\mathrm{s}= \frac{1}{2\gamma} \int_{-D/2}^{+D/2} z\Big[P_{\mathrm {N}}(z)-P_{\mathrm {T}}(z)
\Big]\,\D z \;,
\label{def:zs}
\end{equation}
depends on the different choices made to determine ${\mathbf{P}}^U$. This was already
pointed out by Harasima~\cite{Harasima}.

In Sect.~\ref{sec:results} we want to show for the polymer model considered that the 
differences in $z_\mathrm{s}$ obtained from the IK and H-expressions are small compared to 
the size $\sigma$ of a particle, but not negligible. The ambiguous nature of $z_\mathrm{s}$ 
was discussed in detail in~\cite{RowlinsonWidom,WTR}. In~\cite{WTR} a 
liquid-vapor interface is studied. Since there are no density oscillations near a free 
surface, which are characteristic of liquid-wall interfaces~\cite{Hansen,yoon_kbrev93},
we expect the difference between the IK and H-expressions for $P_{\mathrm{T}}(z)$ 
to be more pronounced for the thin films studied here.
\subsection{The Method of Planes}
\label{subsec:MOP}
Todd, Evans and Daivis~\cite{ToddEvans,ToddEvans2} have introduced a variant of the original
IK-derivation to determine the pressure tensor (termed ``method of planes'')
which avoids the ambiguity of defining a contour to relate two interacting particles. 
The problem is, however, not circumvented because one has to choose a gauge for both 
the pressure tensor and the momentum density~\cite{ToddEvans}. The derivation starts from 
the continuity equations for the mass and momentum and leads to
\begin{eqnarray}
P^{U}_{\alpha z}(z)&\!\!\!=\!\!\!&
\frac{1}{2A} 
\Big \langle \sum^{M}_{i=1} F_{\alpha i}\,\,{\mathrm {sgn}}(z_i-z ) \Big \rangle
\label{eq:ToddEvans:PU:a} \\
&\!\!\!=\!\!\!&
\frac{1}{2A} \Big\langle  \sum_{i\neq j} F_{\alpha ij}
\Big( \Theta(z_i-z)\,\Theta(z-z_j)-\Theta(z_j-z)\,\Theta(z-z_i) \Big) \Big\rangle
\label{eq:ToddEvans:PU:b}
\end{eqnarray}
for the potential part of the pressure tensor and to
\begin{equation}
P^{K}_{\alpha z}(z)=\frac{1}{A} 
\Big \langle \sum_{i=1}^{M} \frac{p_{\alpha i}\,p_{zi}}{m}\,\,\delta(z-z_i) \Big \rangle 
\label{eq:ToddEvans:PK}
\end{equation}
for the kinetic part ($\alpha=x,\, y,\, z$), where $m$ is the mass of a particle. 
In Eq.~(\ref{eq:ToddEvans:PU:a}) ${\mathrm {sgn}} (x)$ is the sign function (= $1$ 
if $x>0$ and $-1$ for $x<0$),
and $F_{\alpha i}$ is the $\alpha$-component of the force exerted on particle $i$
by all other particles. Furthermore, $\Theta(x)$ denotes the Heaviside step function
and $p_{\alpha i}$ is the $\alpha$-component of the momentum of particle $i$.
Using the identity 
\[ 
|z_{ij}|\,\Theta\bigg(\frac{z-z_i}{z_{ij}}\bigg )\,\Theta\bigg(\frac{z_j-z}{z_{ij}}\bigg )
= -z_{ij}\,\bigg[\Theta(z_i-z)\,\Theta(z-z_j)-\Theta(z_j-z)\,\Theta(z-z_i) \bigg] 
\]
one can verify that the diagonal components of the Eqs.~(\ref{eq:ToddEvans:PU:b}) and 
(\ref{eq:ToddEvans:PK}) yield the IK-expression for the normal pressure 
[Eq.~(\ref{eq:p_N:RaoBerne})]. Since Eq.~(\ref{eq:ToddEvans:PU:a}) contains a single
sum instead of the double sum of Eq.~(\ref{eq:p_N:RaoBerne}), it is computationally more
convenient. Therefore, we used Eqs.~(\ref{eq:ToddEvans:PU:a}) and (\ref{eq:ToddEvans:PK})
to calculate the normal pressure. However, these equations are not sufficient
for determining the surface tension $\gamma$, as they do not contain the 
diagonal components of the pressure tensor parallel to the walls, i.e., 
$P_{xx}$ and $P_{yy}$. On the other hand, they
provide a method for the calculation of the viscosity~\cite{ToddEvans}.
\subsection{An approximate formula: IK1-method}
\label{subsec:IK1}
In the literature (see~\cite{Nimajer,Pandey}, for instance) there is still another formula 
for the pressure tensor, which is a kind of a ``tensorized'' version of the Harasima 
expression~(\ref{eq:p_T:Harasima}) (called ``IK1'' in \cite{ToddEvans})
\begin{eqnarray}
{\mathbf P}^{\mathrm{IK1}} (z) &=&\rho(z)k_{\mathrm B}T\;  \hat{\mathbf 1} - 
\frac{1}{2A} \Big \langle \sum_{i\neq j}
\frac{{\mbf{r}}_{ij}{\mbf{r}}_{ij} }{r_{ij}} \,\, U'(r_{ij})
\, \delta(z_{i} - z)\Big \rangle \;.
\label{eq:IK1:film}
\end{eqnarray}
Todd, Evans and Daivis~\cite{ToddEvans} noticed that Eq.~(\ref{eq:IK1:film})
is equivalent to a zeroth-order approximation of the (full) IK-expression and gave
a physical interpretation of the approximation in $k{\mathrm{-space}}$
(see Eq.~(24) in~\cite{ToddEvans}). One can also find a real-space interpretation 
in the following way. If one replaces the integral over $\alpha$ in Eq.~(\ref{eq:IK:A6'})
by the value of the integrand at the lower bound $\alpha=0$, one obtains
\begin{equation}
{\mathbf P}^{U} ( {\mbf{r}} ) = -\frac{1}{2} \Big\langle \sum_{i \neq j}
\frac{{\mbf{r}}_{ij}{\mbf{r}}_{ij} }{r_{ij}} \,\, U'(r_{ij})
\, \delta({\mbf{r}}_{i}-{\mbf{r}})\Big\rangle \;,
\label{eq:IK1:general}
\end{equation}
which gives the potential part of the IK1-expression (\ref{eq:IK1:film}) after
averaging over the tangential coordinates.

Thus, the IK1-method corresponds to the assumption that the two-particle 
density $\rho^{(2)}({\mbf{r}}_{1};{\mbf{r}}_{2})$ is unchanged upon
translation of both arguments along the line ${\mbf{R}}={\mbf{r}}_{2}-{\mbf{r}}_{1}$
which connects the points 1 and 2. However, the breaking of translational 
invariance is one of the basic characteristics of inhomogeneous systems.
The more the system is inhomogeneous, the more the IK1-expression (\ref{eq:IK1:film}) 
for $P_{\mathrm {N}}(z)$ should become inaccurate. On the other hand, integration over $z$ 
yields the same result as the IK- and H-approaches. Therefore, the 
IK1-method leads to the same surface tension $\gamma$, but to a different 
value for $z_\mathrm{s}$ compared to the other two methods.

In Sect.~\ref{sec:results} we show that the IK1-result for $z_\mathrm{s}$ is too
large to allow for an interpretation of $z_\mathrm{s}$ as the effective position of 
the interface, i.e., as the distance of closest approach of a particle to the wall. 
Furthermore, Eq.~(\ref{eq:IK1:film}) leads to strong oscillations of $P_{\mathrm{N}}$ 
in contrast to the condition of mechanical stability which requires a constant profile 
for $P_{\mathrm{N}}$ (see section~\ref{subsec:mechanical_stability}).
\subsection{Mechanical Stability Requires $ {\mathbf P_{\mathbf {N}}}=\mbox{const}$}
\label{subsec:mechanical_stability}
In equilibrium, mechanical stability requires that the gradient of the pressure tensor
vanishes
\begin{equation}
\nabla \cdot \mathbf {P}={\mathbf 0}\; ,
\label{eq:divP=null}
\end{equation}
where ${\mathbf 0}$ denotes the null vector. For a system with planar symmetry, the 
non-diagonal components of $\mathbf {P}$ must also vanish (otherwise shear forces would 
exist) and the lateral components should be identical. So, we have
\begin{equation}
\frac{\partial P_{xx}}{\partial x} \;\mbf{e}_{x}+ 
\frac{\partial P_{yy}}{\partial y} \;\mbf{e}_{y}+ 
\frac{\partial P_{zz}}{\partial z} \;\mbf{e}_{z}=\mbf{0} \quad \mbox{and} \quad
P_{xx}(\mbf{r})=P_{yy}(\mbf{r}) \;.
\label{eq:divP=null:ausgeschrieben}
\end{equation}
Since $\partial P_{xx}/\partial x=0$, $\partial P_{yy}/\partial y=0$ on the one
hand, and $P_{xx}=P_{yy}$ on the other, the lateral components can be functions of $z$ only.
Furthermore, since $\;\partial P_{zz}/\partial z=0$, the normal component of the 
pressure tensor is independent of the distance from the surfaces and must be identical 
to the external pressure $P_{\mathrm{N,ext}}$. This gives
\begin{equation}
P_{\mathrm{N}}(z) = P_{zz}=P_{\mathrm{N,ext}}={\mathrm{const}} \quad \mbox{and} \quad
P_{\mathrm{T}}(z)=P_{xx}(z)=P_{yy}(z) \;,
\label{eq:PN=const}
\end{equation}
i.e., Eq.~(\ref{eq:PnPt}).
The argument presented is not new. It essentially follows the discussion 
of~\cite{RowlinsonWidom} (see p.~44 of~\cite{RowlinsonWidom}). We repeated it here to 
stress the erroneous character of expression~(\ref{eq:IK1:film}). In Sect.~\ref{sec:results}  
we will see that only the IK- (or H-) formula~(\ref{eq:p_N:RaoBerne}) satisfies
condition~(\ref{eq:PN=const}). The independence of Eq.~(\ref{eq:p_N:RaoBerne}) on $z$ was 
already proved analytically in the work of Harasima (see p.~224 of \cite{Harasima}). This 
important properties helps us to set the pressure in the simulations for a given wall 
separation and temperature.
\section{SIMULATION OF POLYMERIC FILMS}
\label{sec:sim}
\subsection{Model}
We study a Lennard-Jones model for a polymer melt~\cite{CB:PRE57} embedded between
two impenetrable walls. All simulation results are given in Lennard-Jones (LJ) units.
Two potentials are used for the interaction between particles. The first one is a 
truncated and shifted LJ-potential which acts between all pair of particles
regardless of whether they are connected or not,
\[
U_{\mbox{\scriptsize LJ-ts}}(r)=\left\{ 
\begin{array}{ll}
U_\mathrm{LJ}(r)-U_\mathrm{LJ}(r_{\mathrm{c}}) & \mbox {if $r<r_{\mathrm c}$}\;, \\
0 & \mbox {otherwise}\;,
\end{array}
\right.
\]
where 
\[
U_\mathrm{LJ}(r)= 4 \epsilon \Big[ (r/\sigma)^{12} -(r/\sigma)^{6}\Big]
\]
and $r_\mathrm{c}=2\times 2^{1/6}$.
The connectivity between adjacent monomers of a chain is ensured by a
FENE-Potential~\cite{KremerGrest}
\[
U_{\mathrm{FENE}}(r)=-\frac{k}{2} R^2_0 \ln \bigg[1-\Big (\frac{r}{R_0}\Big)^2\bigg]
\;,
\]
where $k=30$ is the strength factor and  $R_0=1.5$ the maximum allowed length of a bond.
The wall potential was chosen as
\begin{equation}
U_{\mathrm W}(z)=\bigg( \frac{\sigma}{z}\bigg )^9 \; ,
\label{eq:def:uw}
\end{equation}
where $z=|z_{\mbox{\scriptsize particle}}-z_{\mbox{\scriptsize wall}}|$
($z_{\mbox{\scriptsize wall}}= \pm D/2$). This corresponds to 
an infinitely thick wall made of inifinitely small particles which interact
with inner particles via the potential $180(r/\sigma)^{-12}/(\pi\rho_{\mathrm{wall}})$
where $\rho_{\mathrm{wall}}$ denotes the density of wall particles. The sum 
over the wall particles then yields $(\sigma/z)^{9}$.

The static and dynamic properties of this model were studied in the bulk when
gradually supercooling towards the glass transition \cite{CB:PRE57,bbp,bpbb,bbpb,bdbg}.
The model begins to develop sluggish relaxation if the temperature drops below
$T \approx 0.7$ and yields a critical temperature of mode-coupling theory of
$T_\mathrm{c,bulk} \simeq 0.45$ \cite{bbp} upon further cooling. We quote this value for
the sake of comparison with the film results to be discussed below.
\subsection {Contribution of the Walls to the  Normal  Pressure}
As the wall potential acts only in normal direction, the 
expressions~(\ref{eq:p_T:RaoBerne}) and~(\ref{eq:p_T:Harasima}) for $P_{\mathrm{T}}$ 
remain unchanged. To obtain the contribution of the walls to $P_{\mathrm{N}}$
one can consider each wall as an additional particle of infinite mass
and use Eq.~(\ref{eq:ToddEvans:PU:a}) for the extended system of $M+2$ particles.
Starting from Eq.~(\ref{eq:ToddEvans:PU:a}) one can show that
\begin{eqnarray}
P^{\mathrm{walls,IK}}_{\mathrm{N}}(z) 
&=& \frac{1}{A} \Big\langle \sum_{i=1}^M F_{\mathrm{W}}(z_i-z_{\mathrm{botwall}})\; 
\Theta(z_i-z)\,\Theta(z-z_{\mathrm{botwall}}) \Big\rangle\nonumber \\
&-& \frac{1}{A} \Big\langle \sum_{i=1}^M F_{\mathrm{W}}(z_{\mathrm{topwall}} - z_i)\;
\Theta(z_{\mathrm{topwall}}-z)\,\Theta(z-z_i) \Big\rangle \;,
\label{eq:p_walls:IK}
\end{eqnarray}
where $F_{\mathrm{W}}(z) = -\, {\mathrm{d}}U_{\mathrm{W}}(z)/{\mathrm{d}}z$, 
$z_{\mathrm{botwall}}<z_i< z_{\mathrm{topwall}}$ for all (inner) particles (i.e.,
excluding the wall particles) and $z_{\mathrm{botwall}}<z< z_{\mathrm{topwall}}$ for all 
planes. From Eq.~(\ref{eq:p_walls:IK}) it follows that the force $F_{\mathrm W}$ of a 
wall on a particle contributes to the normal pressure on a given plane if 
the plane lies between the particle and the wall. 

Similarly, one can derive the contribution of the walls within the IK1-approximation 
by starting from Eq.~(\ref{eq:IK1:film}). This yields \cite{mg2000}
\begin{eqnarray}
P^{\mathrm{walls,IK1}}_\mathrm{N}(z) 
&=&  \frac{1}{A}  \Big \langle \sum_{i=1}^M F_{\mathrm{W}}( z_{i} - z_{\mathrm{botwall}})
\;\delta(z_i-z)  \Big \rangle \nonumber \\
&-& \frac{1}{A} \Big \langle \sum_{i=1}^M 
F_{\mathrm{W}}(z_{\mathrm{topwall}} - z_{i})\;  \delta(z_i-z)  \Big \rangle \;,
\label{eq:p_N:IK1:walls}
\end{eqnarray}
where the sum runs over inner particles only, as before. Since $F_{\mathrm{W}}(z_{i} - 
z^\prime)\delta(z_i-z)$ is equivalent to $F_{\mathrm{W}}(z - z^\prime)\delta(z_i-z)$,
$P^{\mathrm{walls,IK1}}_\mathrm{N}(z)$ can be written as a product of the density
profile and a contribution from the walls, i.e., 
\[
P^{\mathrm{walls,IK1}}_\mathrm{N}(z) = \Big[F_{\mathrm{W}}( z - z_{\mathrm{botwall}}) -
F_{\mathrm{W}}(z_{\mathrm{topwall}} - z)\Big ] \rho(z) \;.
\]
\subsection {About the Simulation}
The equilibration of the system was done in the NpT-Ensemble. The production runs, 
however, were performed in the NVT-Ensemble because we are also interested in analyzing
the dynamics of the films later on (for preliminary results see \cite{vbb_confit}). 

At the beginning of the simulation the velocities of all particles were set to zero and 
NRRW- (Non-Reversal-Random-Walk-) chains were ``synthesized'', i.e., only the average 
bond length and bond angle (known from previous bulk simulations) were used to build a 
chain of $N$ ($=10$) monomers. This initial state corresponds to very high energies 
(usually $E(t=0) > 10^{10}$) due to the occurrence of extremely short distances
between non-bonded monomers. 

The surplus of energy must be removed to prevent numerical instabilities. For the bulk 
this can be done by replacing the full LJ-potential by a softer one. The LJ-potential 
is then switched on smoothly~\cite{KremerGrest}. For our model, however, it was necessary 
to keep the (full) \emph{wall} potential from the very beginning of the simulation to 
avoid penetration of the walls. We thus left the potentials unchanged, but used an adaptive
time step: First, the maximum force $\mbf{F}_{\mathrm{max}}$ and the maximum velocity
$v_{\mathrm{max}}$ were determined. A time step $\Delta$ was then chosen so that the
resulting displacement of a particle, which is subject to $\mbf{F}_{\mathrm{max}}$ and 
moves with initial velocity $v_{\mathrm{max}}$ in direction of ${\mbf{F}}_{\mathrm{max}}$,
would be $\mathrm{d}r_\mathrm{max}=10^{-3}$. This (empirical) value is only 
applicable if $\mbf{F}_{\mathrm{max}}$ does not point in direction of a 
bond vector whose size $b$ is closer to the maximum bond
length $R_0$ (see $U_\mathrm{FENE}$) than $10^{-3}$,
since a displacement of this size could break the bond. In such a situation 
we chose $\mathrm{d}r_\mathrm{max}=(R_0 - b_\mathrm{max})/2$ instead of $10^{-3}$
to adjust the time step ($b_\mathrm{max}$ denotes the largest measured bond length).
The equations of motion were then integrated with this time step and the
procedure was repeated. 

After about 250 MD steps the velocities of 
all particles were renewed by drawing them from the Maxwell distribution, and the time
derivative of the volume was set to zero. These steps are important to warrant the numerical
stability of our procedure. Our criterion for the end of this stage was that the minimum 
distance between particles should not be smaller than a certain value, empirically 0.8,
and that the normal pressure of the system should not be too far away from the 
external value, i.e., $|\bar{P}_\mathrm{N}(t)-P_{\mathrm{N,ext}}|/P_{\mathrm{N,ext}} \leq 10^{-2}$,
where $\bar{P}_{\mathrm{N}}(t)$ was computed as an average over the last 20 samples
preceding time $t$. The sample distance was empirically chosen to $10\,\mathrm{exp}(1/T)$
MD steps to take into account stronger correlations at lower temperatures.
Since we kept the film thickness $D$ fixed, 
the simulation at constant pressure 
was realized by varying the area ($=A$) of the simulation box parallel to the walls.
During this initial stage a high bath temperature, $T\!=\!1$, was used.

After this initial stage (with a typical duration of $10^5$ MD steps) the time step could be 
set to $\Delta=0.003$. This value is close to
that used in previous bulk simulations~\cite{CB:PRE57}. The system was then slowly cooled down
to the desired temperature by gradually reducing temperature in a step-wise fashion: 
The bath temperature was set to the next smaller value and the system was propagated for a 
a certain amount of time before the bath temperature was decreased again.

At the end of the cooling process the sampling of the mean-square displacement of the
chain centers parallel to the walls, $g_{3||}(t)$, and of the volume was started. The system 
was propagated until $g_{3||} \geq 9R^2_{\mathrm{ee||}}$ where $R_{\mathrm{ee||}}$ denotes 
the component of the chain's end-to-end vector parallel to the walls. This criterion suffices 
to reach the free diffusive limit and to equilibrate the system completely. During this period 
the system volume was sampled once every 1000 time steps and the average volume of the system 
was calculated. The equilibrated configuration was then further propagated until the 
instantaneous volume reached the average value within a given relative accuracy, usually 
$10^{-5}$. At this point the program fixes this volume and switches to a (pure) 
Nose-Hoover-Algorithm (NVT-Ensemble) for production runs in the canonical ensemble.
During a production run sampling was done once every 1000 time steps.
\section{Results}
\label{sec:results}
\subsection{Profiles of ${\mathbf{P_{\mathbf{N}}(\mbf{z})}}$: IK1 versus (full) IK}
In order to analyze the pressure profiles for our model we studied different
film thicknesses ($D\!=\!3,5,10,20$) at various temperatures
while always keeping $P_\mathrm{N,ext}=1$. For this external pressure many results 
for the bulk behavior are known \cite{CB:PRE57,bbp,bpbb,bbpb,bdbg}. Here, we want
to discuss two representative cases: $D\!=\!3$ ($\approx 2 R_\mathrm{g}$ where $R_\mathrm{g} 
\simeq 1.45$ is the bulk radius of gyration) at $T\!=\!1$, and $D\!=\!10 $ ($\approx 7 R_\mathrm{g}$) 
at $T\!=\!0.42$. The temperature $T\!=\!1$ corresponds to the high-temperature,
(ordinary) liquid state of the melt, whereas $T\!=\!0.42$ belongs to the supercooled
temperature regime close to the critical temperature of mode-coupling theory
($T_\mathrm{c}(D\!=\!10) \approx 0.39$ \cite{vbb_confit}). 

For a film of thickness $D\!=\!3$ ten independent runs of $10^6$ time steps
were simulated at $T\!=\!1$ and $P_{\mathrm{N,ext}}\!=\!1$. The total number of 
particles was 1000 corresponding to 100 chains of length $N=10$ (this number of 
monomers per chain was always kept fixed in our simulations). For $D\!=\!10$ 
five independent runs were done at $T\!=\!0.42$. The length of a run 
was $4.4\times 10^{7}$ time steps. Samples were taken every 1000 steps.
The much longer simulation time in this case is necessary to allow for a 
detailed analysis of the dynamics of the system which is very slow at this temperature.

\begin{figure}[pbt]
\epsfysize=120mm
\epsffile{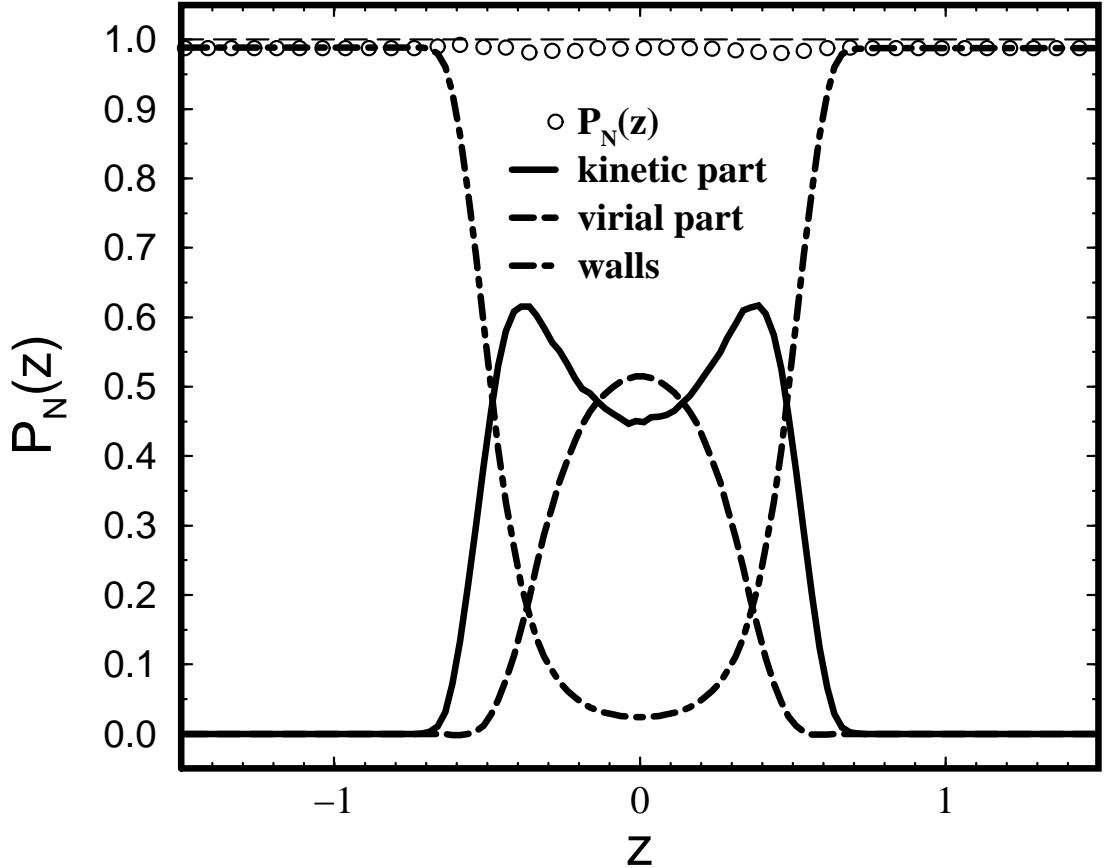}
\caption[]{
Different contributions to the normal pressure profile $P_{\mathrm{N}}(z)$  for a 
film of thickness $D\!=\!3$ ($\approx 2 R_\mathrm{g}$) at $T\!=\!1$ (high-temperature
liquid state) and $P_{\mathrm{N,ext}}\!=\!1$ according to the (full) 
IK-method~[see Eq.~(\ref{eq:p_N:RaoBerne})]. The H-method yields the same result
[see Eq.~(\ref{eq:pN:H})]. The various parts, kinetic (full line), virial (dashed line) 
and wall (dash-dotted), mutually balance one another to yield a constant profile 
$P_\mathrm{N}(z)=P_{\mathrm{N,ext}}$ (circles), as required by mechanical stability 
(see Sect.~\ref{subsec:mechanical_stability}). The difference between $P_{\mathrm{N,ext}}
\!=\!1$ (vertical dashed line) and $P_\mathrm{N}(z)$ shows the accuracy to which we
can fix $P_{\mathrm{N,ext}}$ in the simulation for this film thickness. The difference
is smaller than $2\%$.
}
\label{fig:pz_Anteile_T1_p1_D3_WTR}
\end{figure}

\begin{figure}[pbt]
\epsfysize=120mm
\epsffile{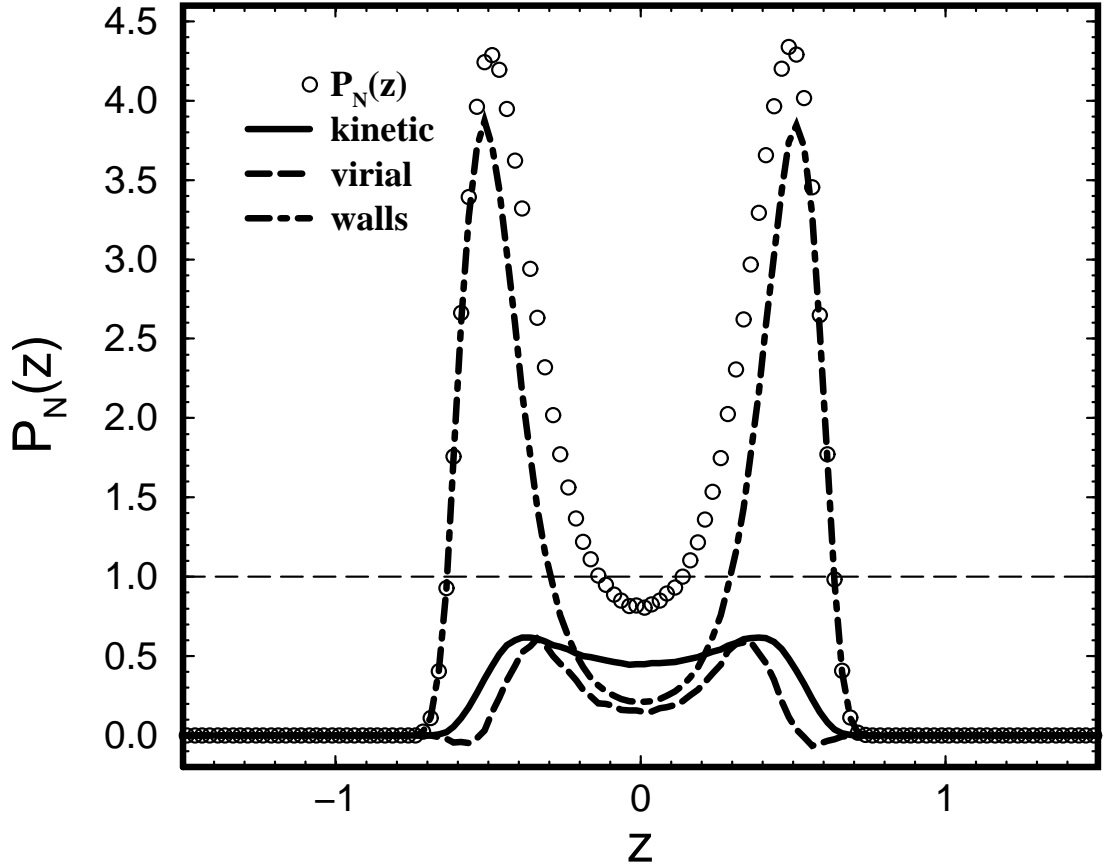}
\caption[]{
Different contributions to the normal pressure profile $P_{\mathrm{N}}(z)$  for a 
film of thickness $D\!=\!3$ ($\approx R_\mathrm{g}$) at $T\!=\!1$ (high-temperature
liquid state) and $P_{\mathrm{N,ext}}\!=\!1$ (vertical dashed line)
according to the IK1-method~[see Eq.~(\ref{eq:IK1:film})]. Contrary to 
Fig.~\ref{fig:pz_Anteile_T1_p1_D3_WTR}, the various parts, kinetic (full line), 
virial (dashed line) and wall (dash-dotted), do not balance, but amplify one 
another, resulting in a (non-physical) oscillatory structure of $P_\mathrm{N}(z)$ 
(circles).
}
\label{fig:pz_Anteile_T1_p1_D3}
\end{figure}

Figures~\ref{fig:pz_Anteile_T1_p1_D3_WTR} and~\ref{fig:pz_Anteile_T1_p1_D3} show
the simulation results for the normal component of the pressure tensor, $P_{\mathrm{N}}$,
calculated according to the IK- and IK1-prescriptions, respectively 
[see Eq.~(\ref{eq:p_N:RaoBerne}) and Eq.~(\ref{eq:IK1:film})]. Furthermore, they 
resolve the 
different contributions stemming from the kinetic part, the virial
(forces between inner particles, i.e. excluding the walls) and the walls.
The striking difference between both prescriptions is that the IK1-method yields
strong oscillations, whereas the pressure profile of the IK-method is constant
throughout the film, in agreement with the condition of mechanical stability
[see Sect.~\ref{subsec:mechanical_stability}]. This deficiency of the IK1-method
has already been pointed in an analysis of the pressure tensor for a simple
liquid \cite{ToddEvans,ToddEvans2}.

\begin{figure}[pbt]
\epsfysize=120mm
\epsffile{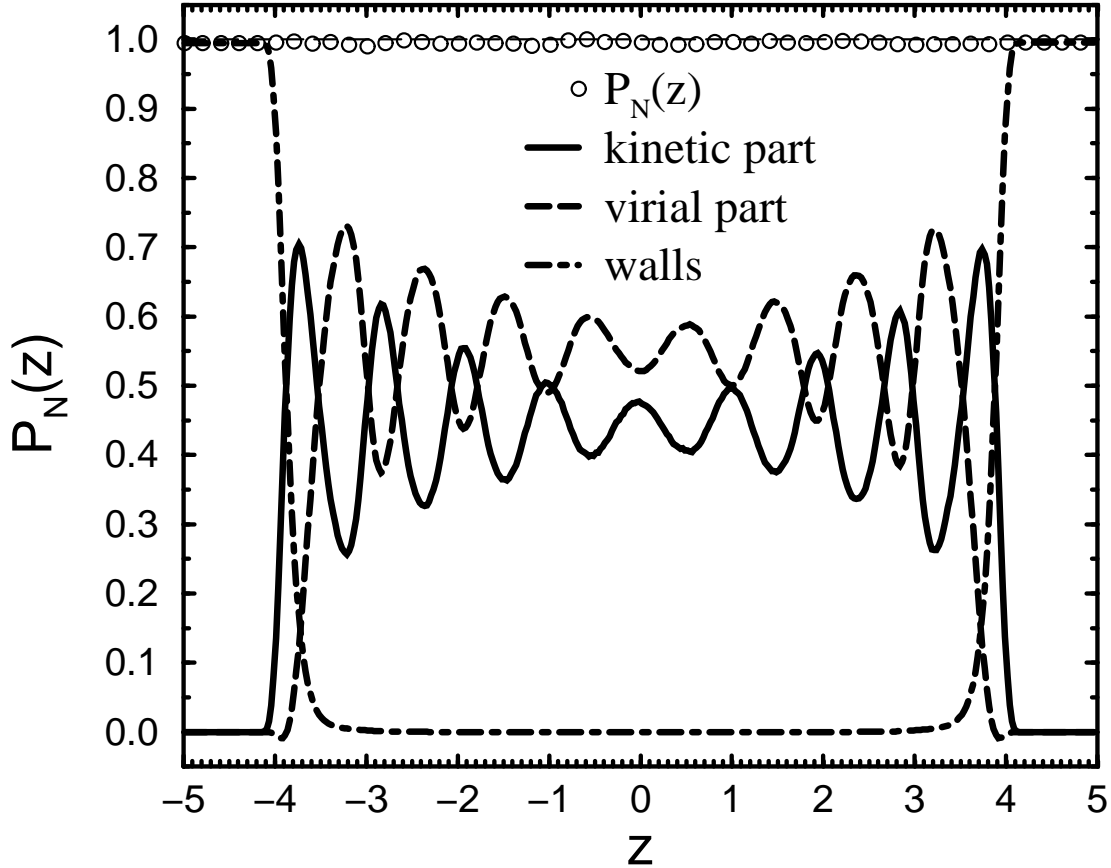}
\caption[]{
Different contributions to the normal pressure profile $P_{\mathrm{N}}(z)$ for a
film of thickness $D\!=\!10$ ($\approx 7 R_\mathrm{g}$) at $T\!=\!0.42$ (supercooled state close
to $T_\mathrm{c} \approx 0.39$ \cite{vbb_confit}) and $P_{\mathrm{N,ext}}\!=\!1$
(vertical dashed line) according to the IK-method~[see Eq.~(\ref{eq:p_N:RaoBerne})]. 
The H-method gives the same result [see Eq.~(\ref{eq:pN:H})]. As in 
Fig.~\ref{fig:pz_Anteile_T1_p1_D3_WTR}, the various parts,
kinetic (full line), virial (dashed line) and wall (dash-dotted), mutually balance
one another and sum up to a constant profile $P_\mathrm{N}(z)=P_{\mathrm{N,ext}}$ (circles),
in agreement with the condition of mechanical stability (see Sect.~\ref{subsec:mechanical_stability}).
}
\label{fig:pz_Anteile_T0.42_p1_D10_WTR}
\end{figure}

Since the kinetic contribution to $P_{\mathrm{N}}$ is proportional to the 
density profile $\rho(z)$, Fig.~\ref{fig:pz_Anteile_T1_p1_D3_WTR} shows
that practically no particle is present in the vicinity of the walls.
The excluded-volume interaction creates a depletion zone of about $0.8$ 
between the wall ($z_\mathrm{wall} = \pm 1.5$) and the monomer positions at this temperature.
Any plane in this region separates all particles of the system, which lie
on the side of the plane facing towards the inner part of the film, from the wall 
on the other side. There is no interparticle force across the plane and 
thus the virial contribution to the normal pressure vanishes. The behavior 
of $P_{\mathrm{N}}(z)$ near the wall arises only from the wall-particle interaction.
This interaction does not 
depend on the position of the plane as long as all the particles stay on the 
opposite side, i.e., as long as $\rho(z)\!\approx \!0$. This explains why 
$P_{\mathrm{N}}$ is constant in the region close to the walls. With increasing 
distance from the wall the density starts to increase from zero. Then, the kinetic 
and virial parts begin to contribute, whereas the effect of the walls decreases.
In this intermediate region none of the contributions is negligible, but their sum
still remains constant, in accord with Eq.~(\ref{eq:PN=const}). Very far from 
the walls the contribution of the walls to $P_{\mathrm{N}}$ becomes negligible.
There, one expects that the variations of the kinetic and virial terms must be 
opposite to each other. A first indication of this opposite behavior can be observed in
Fig.~\ref{fig:pz_Anteile_T1_p1_D3_WTR}. A better demonstration is, however, shown in 
Fig.~\ref{fig:pz_Anteile_T0.42_p1_D10_WTR} where the film thickness
is large enough to exhibit an inner region with negligible wall contribution.

\begin{figure}[pbt]
\epsfysize=120mm
\epsffile{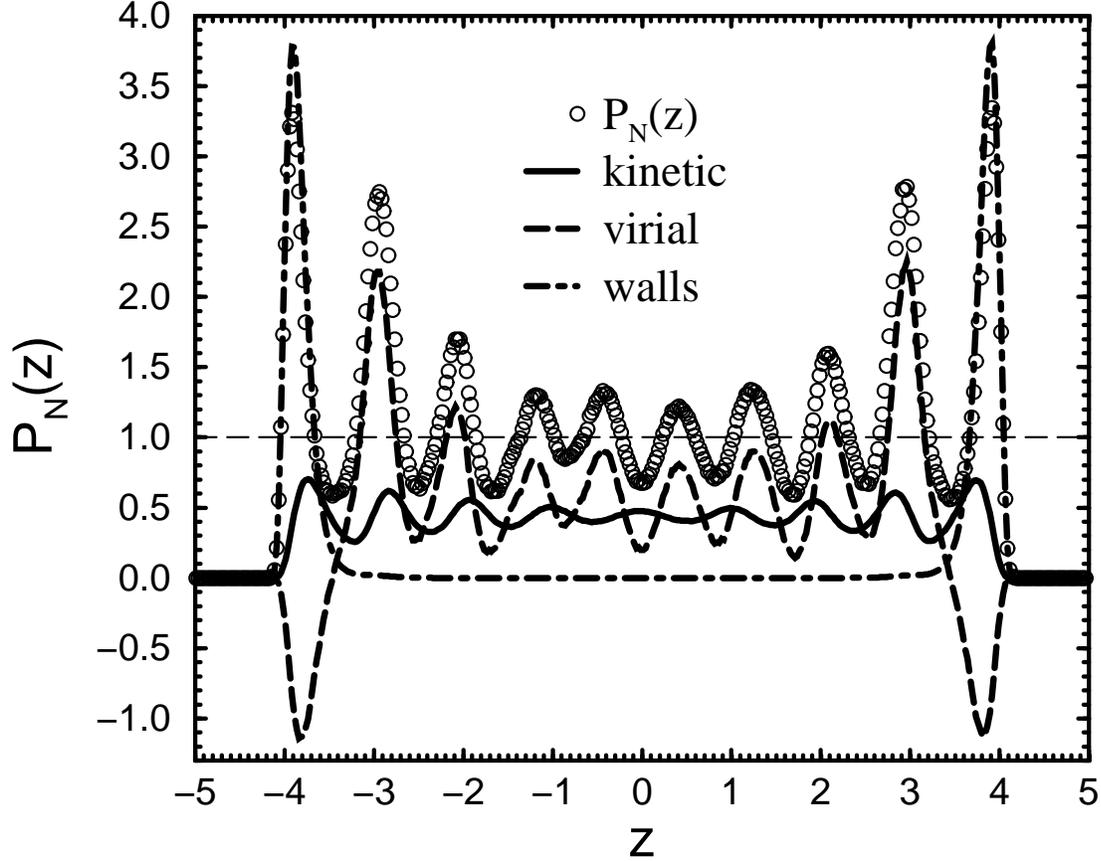}
\caption[]{
Different contributions to the normal pressure profile $P_{\mathrm{N}}(z)$ for a 
film of thickness $D\!=\!10$ ($\approx 7 R_\mathrm{g}$) at $T\!=\!0.42$ (supercooled state close
to $T_\mathrm{c} \approx 0.39$ \cite{vbb_confit}) and $P_{\mathrm{N,ext}}\!=\!1$
(vertical dashed line) according to the IK1-method~[see Eq.~(\ref{eq:IK1:film})]. As in 
Fig.~\ref{fig:pz_Anteile_T1_p1_D3}, the various parts, kinetic (full line), 
virial (dashed line) and wall (dash-dotted), give rise to a non-constant 
pressure profile (circles) contrary to the requirement of mechanical stability.
}
\label{fig:pz_Anteile_T0.42_p1_D10}
\end{figure}

Contrary to that, the various contributions of the IK1-methods are (almost) in
phase. Figure~\ref{fig:pz_Anteile_T1_p1_D3} illustrates that the strong deviation 
of $P_{\mathrm{N}}^{\mathrm{IK1}}$ from a constant is caused by the interaction of
the wall with the monomers close to the maximum of $\rho(z)$ if $D\!=\!3$.
If the film thickness increases, Fig.~\ref{fig:pz_Anteile_T0.42_p1_D10} shows that the
oscillations of $P_{\mathrm{N}}$ propagate through the whole film. Close to the wall, the 
dominant contribution still comes from the wall-monomer interaction, whereas the oscillations
in the inner part of the film are in phase with the virial. The contribution of the virial
is negative close to the wall, reflecting a predominantly attractive interaction between the 
monomers. This dominance of the attractive interaction is also visible for the (correct)
IK-method, but is much less pronounced in this case.

\begin{figure}[pbt]
\epsfysize=120mm
\epsffile{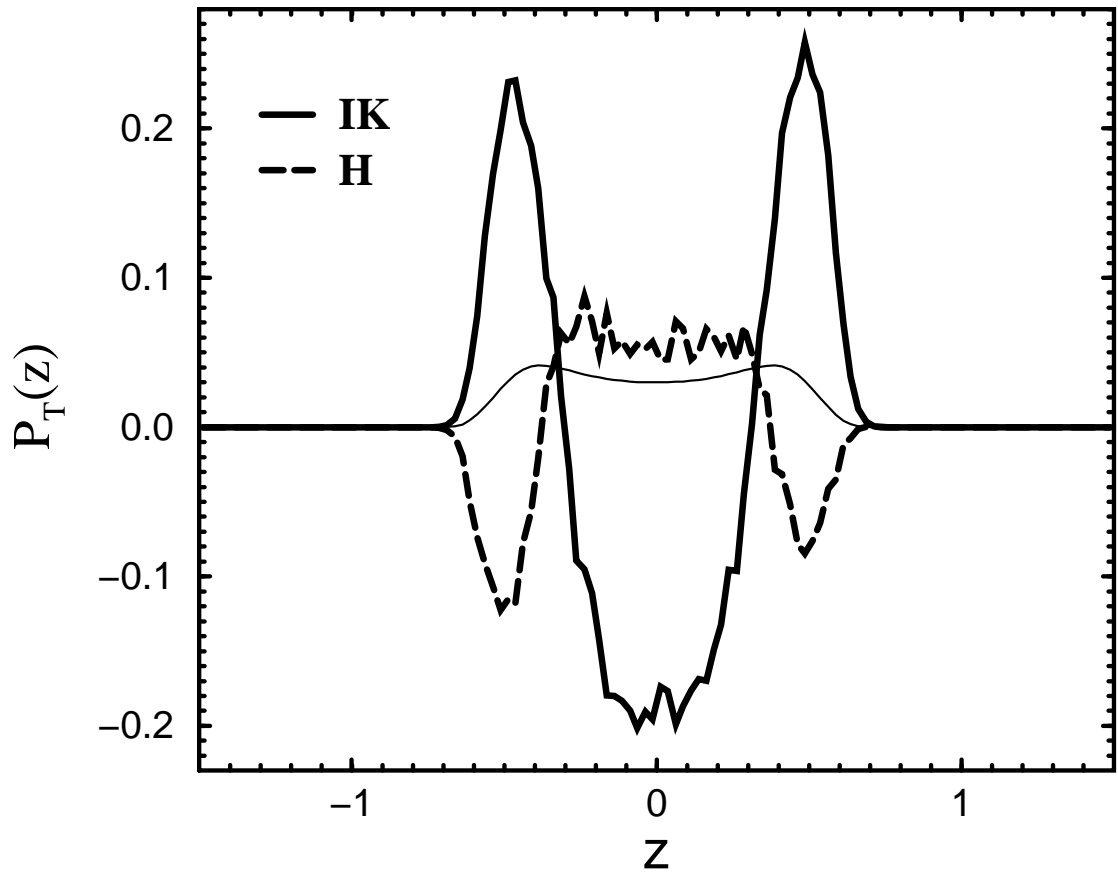}
\caption[]{
Tangential component $P_{\mathrm{T}}(z)$ of the pressure tensor as obtained from the 
IK-for\-mu\-la [Eq.~(\ref{eq:p_T:RaoBerne})] and from the H-formula [Eq.~(\ref{eq:p_T:Harasima})] 
for $D\!=\!3$ ($\approx 2 R_\mathrm{g}$), $T\!=\!1$ (high-temperature liquid state) 
and $P_{\mathrm{N,ext}}\!=\!1$. The thin solid line shows the kinetic contribution
$k_\mathrm{B}T\rho(z)$ (divided by 15 to put it on the scale of the figure).
}
\label{fig:pxy_comp_T1_p1_D3}
\end{figure}

\begin{figure}[pbt]
\epsfysize=120mm
\epsffile{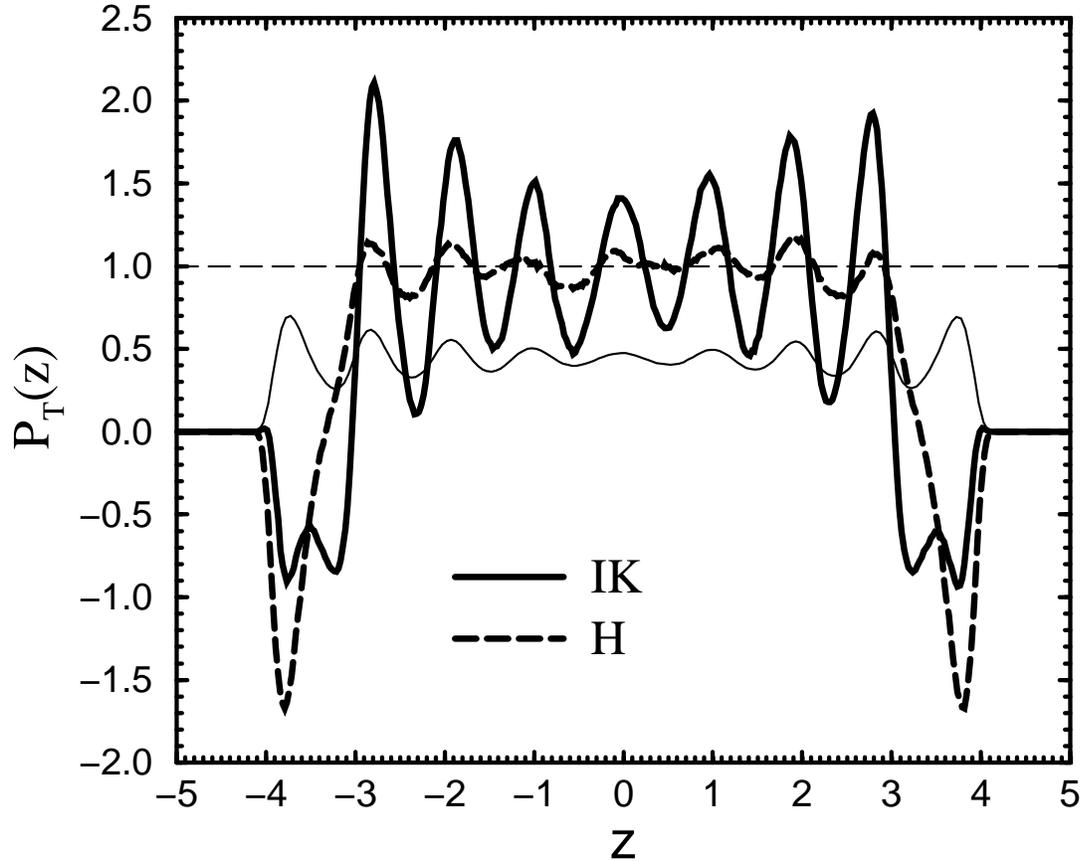}
\caption[]{
Tangential component $P_{\mathrm{T}}(z)$ of the pressure tensor as obtained from the 
IK-for\-mu\-la [Eq.~(\ref{eq:p_T:RaoBerne})] and from the 
H-formula [Eq.~(\ref{eq:p_T:Harasima})] for $D\!=\!10$ ($\approx 7 R_\mathrm{g}$), 
$T\!=\!0.42$ (supercooled state close to $T_\mathrm{c} \approx 0.39$ \cite{vbb_confit})
and $P_{\mathrm{N,ext}}\!=\!1$ (vertical dashed line).  The thin solid line shows the 
kinetic contribution $k_\mathrm{B}T\rho(z)$.
}
\label{fig:pxy_comp_T0.42_p1_D10}
\end{figure}

The situation becomes more complicated when studying the lateral component of the pressure
tensor. Here, the two alternative formulas, Eqs.~(\ref{eq:p_T:RaoBerne}) and 
(\ref{eq:p_T:Harasima}), can yield completely different profiles.
Figures~\ref{fig:pxy_comp_T1_p1_D3} and~\ref{fig:pxy_comp_T0.42_p1_D10} compare
the IK and the H-versions to calculate the lateral pressure $P_{\mathrm {T}}(z)$ 
for $D\!=\!3$, $T\!=\!1$ and $D\!=\!10$, $T\!=\!0.42$, respectively. Whereas both methods oscillate in 
phase with one another for the thicker film, they are anti-correlated for $D\!=\!3$. The
lateral pressure of the IK-method is positive close to the walls, but negative in the
middle of the film, whereas the behavior is just vice versa for the H-method. Due to the
aforementionend ambiguity of $P_{\mathrm {T}}(z)$ it is impossible to
decide which methods yields the physically more realistic result. If the film thickness
increases, the qualitative difference between the IK- and H-methods (almost) vanishes
and only quantitative differences remain. The oscillations of $P_{\mathrm {T}}(z)$
clearly reflect the monomer profile. In the inner portion of the film they are much weaker 
for the H-method than for of the IK-method. This is related to the local nature of 
Eq.~(\ref{eq:p_T:Harasima}) due to the presence of delta-function. Density oscillations 
are thus incorporated not only in the kinetic term, but also also in the virial part of 
the Harasima formula. Both terms partially cancel each other. Although the profile
generated by Eq.~(\ref{eq:p_T:Harasima}) is thus closer to $P_\mathrm{N,ext}$ than 
that of Eq.~(\ref{eq:p_T:RaoBerne}), this should not be considered as an argument in favor 
of the H-method. 
A clear distinction between both methods would only be possible if one could find a
quantity which specifically probes $P_{\mathrm {T}}(z)$ and whose behavior is known
{\em a priori}, as it was the case for $P_{\mathrm {N}}(z)$.

\subsection{Surface Tension and Surface of Tension}
\begin{figure}[pbt]
\epsfysize=120mm
\epsffile{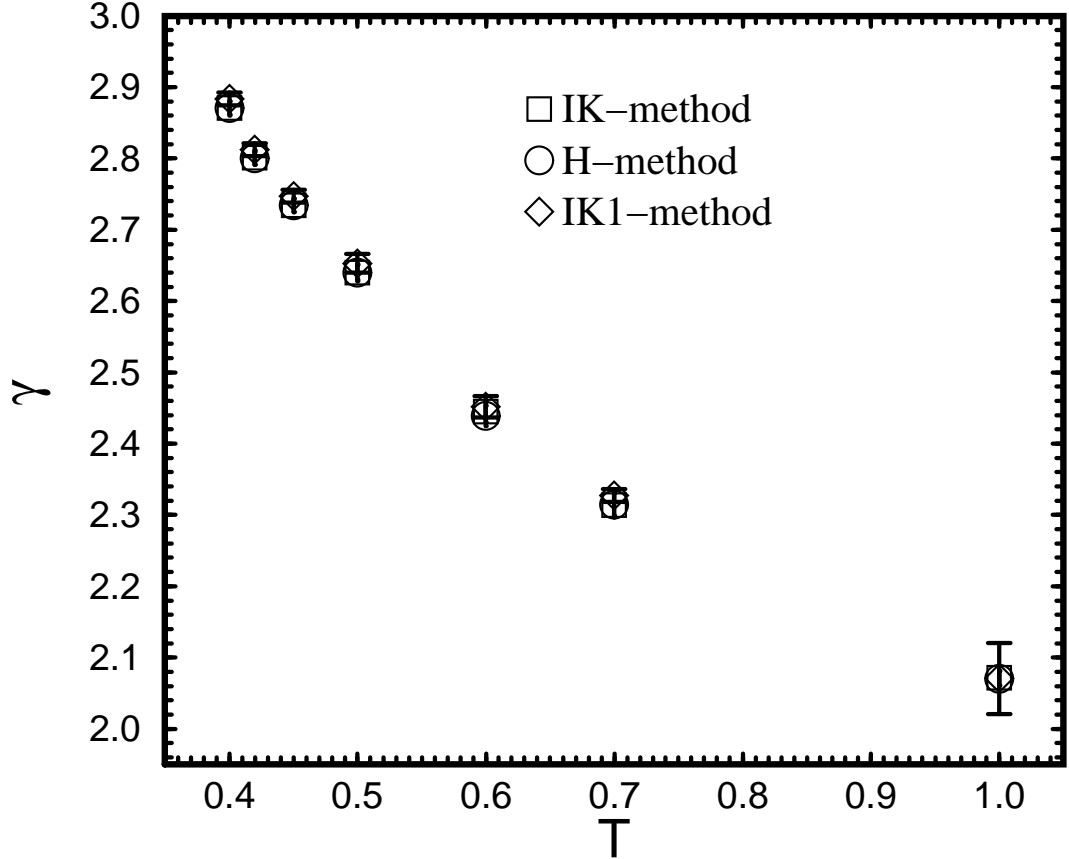}
\caption[]{
Temperature dependence of the surface tension, $\gamma$, calculated by 
Eq.~(\ref{eq:def:gamma}), using the IK-, H- and IK1-methods for $D\!=\!5$ ($\approx
3 R_\mathrm{g}$) and $P_{\mathrm{N,ext}}\!=\!1$. The temperatures shown range from the 
high-temperature, liquid state of the film to the supercooled state. 
}
\label{fig:gammaD5}
\end{figure}

As mentioned in Sect.~\ref{subsec:ikh}, integration of the pressure profiles over $z$ 
yields the same result for the IK-, H- and IK1-expressions. Therefore, all methods
must lead to the same surface tension $\gamma$ [i.e., Eq.~(\ref{eq:kb:gamma})]. This 
expectation is nicely borne out by the simulation data for all film thicknesses and 
temperatures studied, where $\gamma$ was calculated by Eq.~(\ref{eq:def:gamma}).  
Figure~\ref{fig:gammaD5} exemplifies this behavior for $D\!=\!5$ 
($\approx 3 R_\mathrm{g}$). With decreasing temperature the surface tension increases
by about a factor of 1.5. 

\begin{figure}[pbt]
\epsfysize=120mm
\epsffile{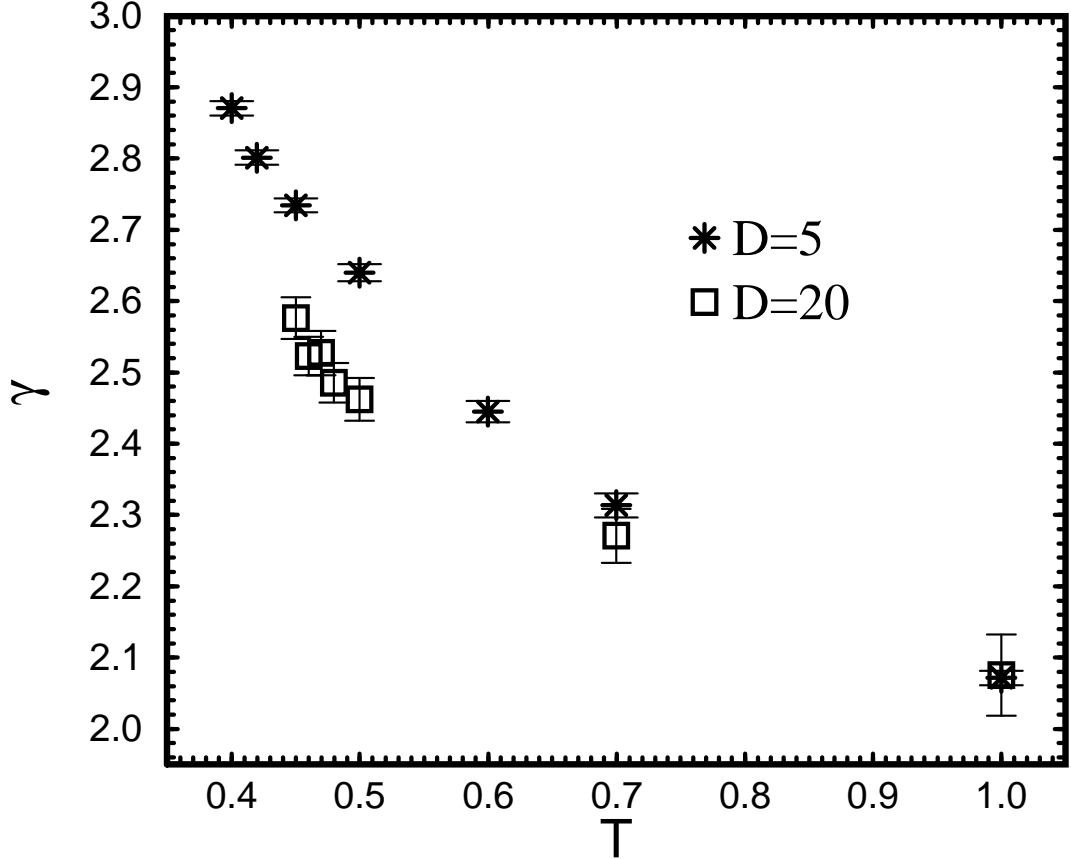}
\caption[]{
Comparison of the temperature dependence of the surface tension, $\gamma$, for
$D\!=\!5$ ($\approx 3 R_\mathrm{g}$) and $D\!=\!20$ ($\approx 14 R_\mathrm{g}$). The 
results of the IK-method are shown only. The other methods (H- and IK1-methods) yield the 
same $\gamma$'s within the error bars, as exemplified in Fig.~\ref{fig:gammaD5} for $D\!=\!5$.
The external pressure is $P_{\mathrm{N,ext}}\!=\!1$. The temperatures shown range from the 
high-temperature, liquid state of the film to the supercooled state.
}
\label{fig:gamma_von_T_D5+20_on_the_fly}
\end{figure}

\begin{figure}[pbt]
\epsfysize=120mm
\epsffile{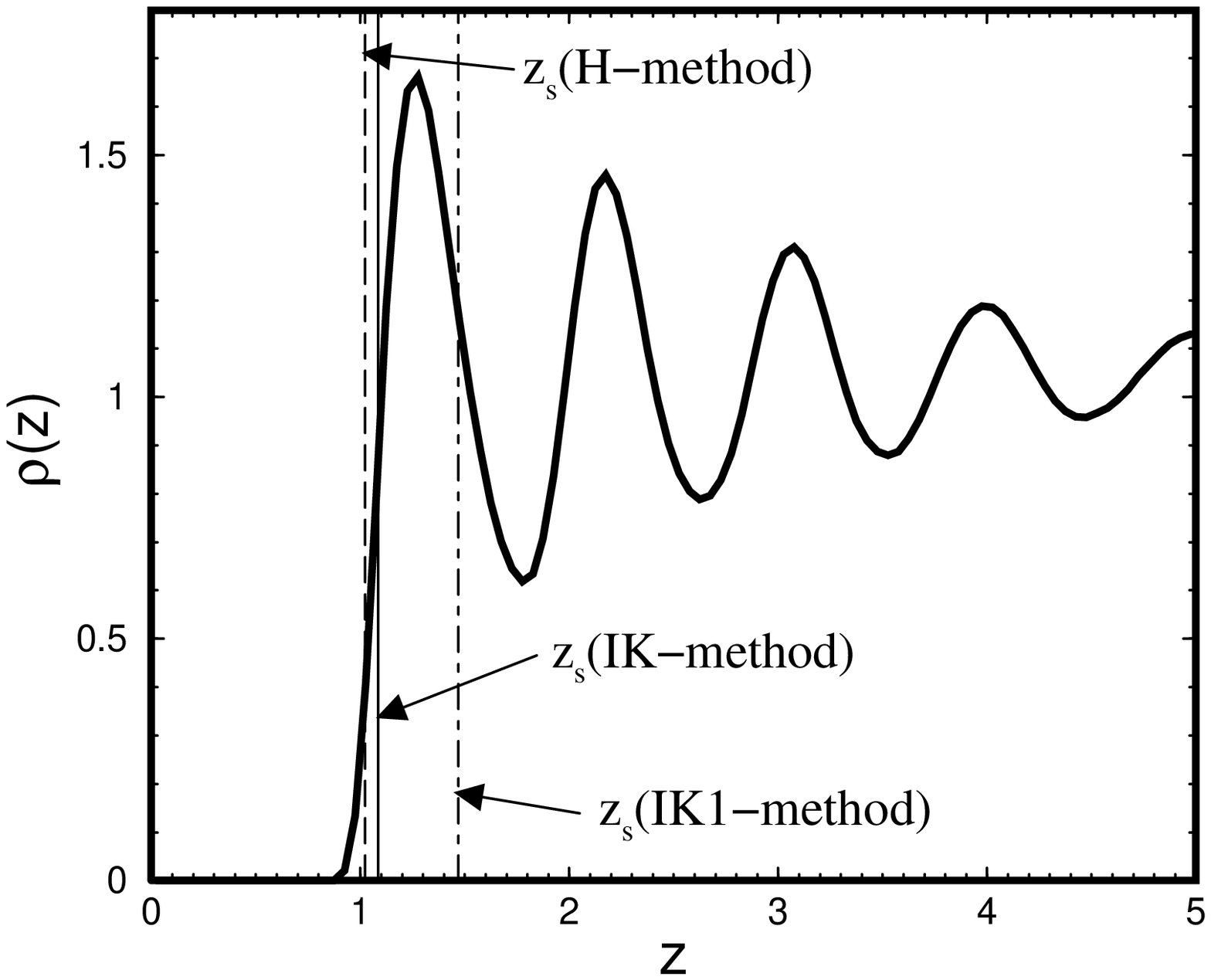}
\caption[]{
Monomer density profile of a film of thickness $D\!=\!10$ ($\approx 7 R_\mathrm{g}$) at 
$T\!=\!0.42$ ($> T_\mathrm{c} \approx 0.39$ \cite{vbb_confit}) and $P_{\mathrm{N,ext}}\!=\!1$. 
Since the profile is symmetric around the middle of the film, the figure only shows one half 
of it. The scale of the abcissa was shifted so that the wall is placed at $z\!=\!0$. The 
vertical lines mark the values of $z_\mathrm{s}$ computed according to the IK-, H- and 
IK1-methods.
}
\label{fig:density_profile_und_zs_comp_T0.42_p1_D10_sym}
\end{figure}

Qualitatively, this temperature dependence is expected. The monomer density of a polymer 
melt close to a hard wall exhibits a profile that is large at the wall and decays 
towards the bulk value in an oscillatory fashion with increasing distance from the wall 
(see Fig.~\ref{fig:density_profile_und_zs_comp_T0.42_p1_D10_sym} as an example)
\cite{yoon_kbrev93}. Since the average density grows with decreasing temperature in 
a simulation at constant pressure, the maxima and minima of the profile become more
pronounced. This means that there are more monomers in the highly populated layers at 
low than at high temperatures, and that the oscillations of profile become more 
long-ranged. These effects tighten the film so that the free energy needed to move 
monomers out of the interface, i.e., the surface tension, should increase as
temperature decreases. The same effect is expected when reducing the film thickness 
because the layering is more pronounced in thinner films. This expectation is borne out 
by the simulation data (see Fig.~\ref{fig:gamma_von_T_D5+20_on_the_fly}).

\begin{figure}[pbt]
\epsfysize=120mm
\epsffile{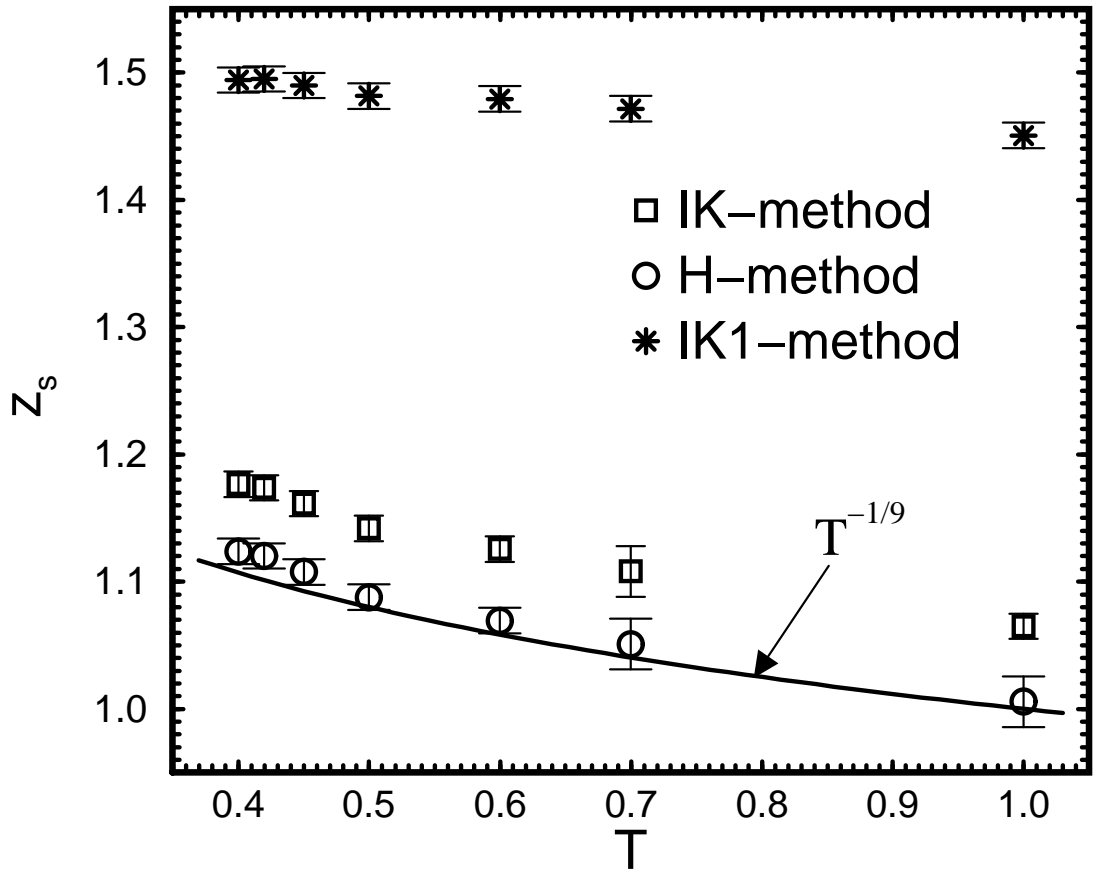}
\caption[]{
Temperature dependence of the surface of tension $z_\mathrm{s}$ [Eq.~(\ref{def:zs})]
determined by the IK-, H- and IK1-methods for $D\!=\!5$ and $P_{\mathrm{N,ext}}\!=\!1$.
The solid line shows the simple estimate, $z_\mathrm{w}=1/T^{1/9}$ [Eq.~(\ref{eq:def:zw})], for
the position of the wall.
}
\label{fig:zsD5}
\end{figure}

Contrary to $\gamma$, the discussion of Sect.~\ref{subsec:ikh} implies that the surface
of tension, $z_\mathrm{s}$, depends on the method applied. This fact is illustrated
in Fig.~\ref{fig:zsD5} which shows the temperature dependence of $z_\mathrm{s}$ for
the IK-, H- and IK1-methods. The difference between IK and the H-methods is rather small,
whereas the IK1-result lies substantially above the values of the other two methods. 
Since $z_\mathrm{s}$ can be interpreted as the distance of closest approach of a 
monomer to the wall, i.e., as the effective position of the wall, the following simple 
argument rules out the IK1-result: At temperature $T$, a particle can only penetrate
into a (soft) wall up to the point, $z_\mathrm{w}$, where the wall potential balances 
thermal energy of the particle, i.e., $U_\mathrm{w}(z_\mathrm{w})/T\!=\!1$. Using 
Eq.~(\ref{eq:def:uw}) this gives
\begin{equation}
z_\mathrm{w} = \bigg(\frac{1}{T}\bigg)^{1/9} \;.
\label{eq:def:zw}
\end{equation}
Equation~(\ref{eq:def:zw}) is compatible with the IK- and H-predictions, but not with
the IK1-result. Another way to illustrate this point is shown in
Fig.~\ref{fig:density_profile_und_zs_comp_T0.42_p1_D10_sym} where we plotted 
the monomer density profile of a film of thickness $D\!=\!10$ at $T\!=\!0.42$.
With increasing film thickness the IK- and H-values for $z_\mathrm{w}$ approach
one another -- for $D\!=\!20$ for example, they are indistinguishable within the error 
bars (not shown here) --, but
 the disparity to the IK1-result remains. The figure clearly shows that the IK1-method
places the effective wall position deeply into the interior of the film, whereas it
has to be situated in the region where the density profile approaches zero.
\section{Conclusions}
We have reported simulation results for the pressure tensor of polymeric thin films 
which investigate the ambiguity in the definition of the potential part of this quantity. 
We studied three common methods: the method of Irving and Kirkwood~\cite{IK}, that of 
Harasima~\cite{Harasima} and an approximation of the IK-method, the so-called 
IK1-approach~\cite{ToddEvans}. On a 
microscopic scale, our simulation results show significant differences between the IK- and 
H-methods for the lateral component $P_{\mathrm{T}}(z)$ of the pressure tensor. However, 
both methods agree with each other for the normal component $P_{\mathrm{N}}(z)$. 
They lead to a constant profile in accord with mechanical stability. On the other hand, 
the IK1-formula exhibits strong oscillations of $P_{\mathrm{N}}(z)$, as also found in
\cite{ToddEvans,ToddEvans2}. The origin of 
this discrepancy comes from the fact that IK1-method corresponds to a zeroth-order approximation 
of the IK-expression, which assumes translational invariance of the two-particle density 
$\rho^{(2)}({\mathbf r}_{1};{\mathbf r}_{2})$ with respect to the difference vector 
$\mbf{R}=\mbf{r}_{2}-\mbf{r}_{1}$. This assumption is not valid in thin films which 
exhibit density oscillations that are damped out only gradually with increasing distance from
the wall. This local structure becomes more pronounced with decreasing temperature and film
thickness. The more pronounced it is, the stronger the IK1-method will deviate from the IK-expression.

However, when integrated over the whole system all methods give the same result.
Thus, the surface tension, $\gamma$, of a planar system can still be calculated
using each of these methods. This is no longer possible for moments of the pressure
profiles, such as the surface of tension $z_\mathrm{s}$. 
The fact that IK1-expression can be used to calculate the surface tension
although it is based on an incorrect expression for the local
pressure tensor has occasionally caused confusion in the literature.
For instance, Pandey et al.~\cite{Pandey} applied the IK1-expressions to
polymer films confined between one repulsive and one attractive wall, taking 
the local pressure profiles literally. The present analysis shows that the 
pressure profiles published in~\cite{Pandey} are incorrect. Thus we hope that 
the present analysis will help to avoid this confusion in future simulation studies.

\bigskip

\noindent{\bf Acknowledgments}\\[2mm]
\noindent
We thank M. M\"uller and J. Horbach for enlightening discussions, and the unknown referee
for valuable comments on the manuscript. Generous grants of simulation 
time by the computer center at the university of Mainz, the HLRZ J\"ulich and the RHRK in 
Kaiserslautern are gratefully acknowledged. This work was supported by the Deutsche 
Forschungsgemeinschaft under SFB262/D2. J.B. is indebted to the European Science 
Foundation for financial support by the ESF Programme on ``Experimental and 
Theoretical Investigations of Complex Polymer Structures'' (SUPERNET).
%
%
      
%
%
\newpage

\begin{thebibliography}{99}
\bibitem{pt1_of_referee}
Strictly speaking, there is also an ambiguity associated with the kinetic contribution 
to the pressure tensor. As pointed out, this contribution can be obtained by
calculating the momentum flux across an infinitesimal surface $\mathrm{d}\mbf{A}$ at a
given time, and then by averaging over time. The question is how the point in time
when the particle crosses $\mathrm{d}\mbf{A}$ is determined? Is it when the 
particle just touches the surface? Or is it when its center of mass coincides
with the position of the surface? Obviously, there are infinitely many
possible definitions of a crossover time. However, all of them should yield
the same result for ${\mathbf{P}}^{K}(\mbf{r})$ when averaging over time
\cite{ToddEvans}. In thermal equilibrium this result is given by
Eq.~(\ref{eq:kin}).
\bibitem{footnote}
Irving and Kirkwood discuss the stress tensor $\mbf{\sigma}^U(\mbf{r})$ which
is equal to $-{\mathbf{P}}^U(\mbf{r})$.
\end{thebibliography}
\end{document}